\documentclass[draft,grl]{agutex2015}

  \usepackage{graphicx}
  \usepackage{color}
  \usepackage{amssymb}
  \usepackage{caption}
  \usepackage{wasysym}

\authorrunninghead{GENESTRETI ET AL.}

\titlerunninghead{PLASMASPHERE TEMPERATURE AND DENSITY}

\begin{document}

%
%

\title{A nearly continuous observation of the equatorial plasmasphere from the inner radiation belt to near a magnetopause reconnection site}

%
%

%
%

\authors{K. J. Genestreti,\altaffilmark{1}\thanks{Formerly at Southwest Research Institute, San Antonio, TX, USA}
	      S. A. Fuselier, \altaffilmark{2,3}
	      J. Goldstein, \altaffilmark{2,3}
	      J. C. Foster, \altaffilmark{4}
	      D. Malaspina, \altaffilmark{5}
	      S. K. Vines, \altaffilmark{6}\thanks{Formerly at Southwest Research Institute, San Antonio, TX, USA}
	      R. Nakamura, \altaffilmark{7}
	      J. L. Burch \altaffilmark{2}}

\altaffiltext{1}{Space Science Center, University of New Hampshire, Durham, NH, USA}	      
\altaffiltext{2}{Space Science and Engineering Division, Southwest Research Institute, San Antonio, TX, USA}
\altaffiltext{3}{Department of Physics and Astronomy, University of Texas San Antonio, San Antonio, TX, USA}
\altaffiltext{4}{Haystack Observatory, Massachusetts Institute of Technology, Westford, MA, USA}
\altaffiltext{5}{Laboratory for Atmospheric and Space Physics, University of Colorado Boulder, Boulder, CO, USA}
\altaffiltext{6}{Applied Physics Laboratory, Johns Hopkins University, Laurel, MD, USA}
\altaffiltext{7}{Space Research Institute, Austrian Academy of Sciences, Graz, Austria}









%
%


\keypoints{\item{We report a nearly continuous observation of the equatorial plasmasphere and plume by VAP and MMS}
	         \item{The proton temperature increases by a factor of $\sim$100 from the inner to outermost extent}
	         \item{The density scales by $\sim L^{-4}$ and decreases by a factor of $\sim$1000 times from the inner to outermost point of observation}}


%
%


\begin{abstract}
On 22 October 2015, VAP and MMS obtained near-continuous observations of the full radial extent of the duskside equatorial plasmasphere and plume.  The plume is evident in in situ plasma data and an equatorial mapping of the ionospheric total electron content. The properties of the equatorial plasmasphere change dramatically from its the inner radiation belt to its outermost boundary (the magnetopause, near a reconnection site). The density decreases by a factor of $\sim$1000 over this range and scales with $L$-shell as $L^{-4.3\pm0.4}$, in good agreement with with theoretical expectations of the expansion of a flux tube volume during outward radial transport. The proton temperature increases by a factor of $\sim$100 over this same range, with the most pronounced heating occurring at $L>7$, which was covered by the orbit of MMS. 
\end{abstract}

%
%

%

\begin{article}

%
%

\section{Introduction}

The plasmasphere is the cold and dense extension of the upper ionosphere into the magnetosphere. The plasmasphere is distinguished from other coupled and often overlapping magnetospheric plasmas by its high density (100s to 10,000s cm$^{-3}$) and low temperature (0.1s to 10s eV).  Owing to its low temperature, plasmaspheric dynamics are governed almost entirely by $\vec{E}\times\vec{B}$ drift. The main plasmasphere torus is typically contained within closed $\vec{E}\times\vec{B}$-drift paths at lower L-shells \cite{Grebowsky.1970,LemaireandGringauz.1998}.

When the solar wind magnetic field points southward, magnetic reconnection allows the solar wind electric field to fractionally penetrate the magnetosphere, opening previously closed $\vec{E}\times\vec{B}$-drift paths and redirecting the previously trapped plasma sunward, forming a plume \cite{Grebowsky.1970,Sandel.2001,GoldsteinandSandel.2005}. The plasmaspheric plume is frequently observed at the noon-to-duskside reconnecting magnetopause [e.g., \textit{Walsh et al.}, 2014], though the impact on reconnection of this dense and cold plasma remains uncertain. Unresolved topics related to the impact of the plume on dayside reconnection include, but are not limited to: modification (or not) of (1) the local reconnection rate \cite{Borovsky.2008,Borovsky.2013,Wang.2015,Fuselier.2017}, (2) the global reconnection rate \cite{Lopez.2010}, (3) both the local and global reconnection rates \cite{Zhang.2016}, (4) the local fields geometry \cite{Malakit.2013}, diffusive scale sizes, and redistribution of magnetic to thermal energy \cite{Wang.2014,ToledoRedondo.2015,ToledoRedondo.2016a,ToledoRedondo.2016b}, and/or (5) downstream exhaust speed \cite{Walsh.2013,Walsh.2014,Fuselier.2017}.

Scaling relations between density, temperature, and geocentric distance reveal that the plasmaspheric plume is warmer and less dense than the main plasmasphere torus. \cite{Chappell.1974} first established an $L^{-4}$ theoretical scaling law to describe the evolution of the density within a lossless flux tube radially transported in a dipole field. Statistical studies have since used this relation to discriminate between the plasmaspheric plume and the trough \cite{Sheeley.2001,Walsh.2014}. Within the plasmasphere main torus, the density and temperature are inversely related \cite{Comfort.1985,Comfort.1986,Comfort.1996}. \cite{Genestreti.2016} found a positive correlation between the plasmaspheric proton temperature and the density of ring current ions within the plasmasphere main torus, which they tentatively attributed to wave particle heating. Based on statistical analysis of Time History of Events and Macroscale Interactions during Substorms (THEMIS) data, \cite{Lee.2014} found that plasmasphere-like ions at large radial distances ($R>5$ $R_\mathrm{E}$) were generally hotter in the afternoon sector than elsewhere, which they attributed to heating within the plume. Additional observations of plume ions at the magnetopause have showed that the plume is significantly hotter and more energetic than the plasmasphere main torus.

In this study, we derive scaling relations between the density, temperature, and $L$-shell location of the plasmasphere. Unlike the aforementioned studies, we derive these scaling relations for one single event, 22 October 2015, when NASA's Van Allen Probes (VAP) and Magnetospheric Multiscale (MMS) missions provided simultaneous and nearly continuous coverage of the full radial extent of the equatorial duskside plasmasphere and plume. Data from one of the several MMS magnetopause encounters on 22 October 2015 have been used to examine the micro-scale influence of the cold plume ions on dayside reconnection \cite{ToledoRedondo.2016b}. For the first time, we use in situ data from a single event to track the temperature and density of the equatorial plasmasphere from the proton radiation belt, to its outermost extent, the magnetopause.

In the following section, we describe the data used in this study and the means by which we determine densities and temperatures. In Section 3, we detail the state of the plasmasphere during the 22 October 2015 conjunction event and derive scaling relations, which show a factor of 1000 decrease in the density and a factor of 100 increase in the temperature from the plasmapause near $L\approx2$ to the dayside reconnection site. We also find that the density scales as $L^{(-4.3\pm0.4)}$, which is within error bars of the theoretical scaling relation of \cite{Chappell.1974}. In the final section, we summarize our results.

\section{VAP and MMS data}

VAP consists of two probes, one leading and one trailing, which have apogees at nearly identical MLT and geocentric distances of 5.8 $R_\mathrm{E}$, with roughly 2 hours of orbital phase difference \cite{Mauk.2013}. MMS is a four spacecraft constellation with each probe separated by tens of km \cite{Burch.2015}. During the first phase of the mission, MMS was in an inclined equatorial orbit with an apogee and perigee at geocentric distances of 12 and 1.1 $R_\mathrm{E}$, respectively \cite{Fuselier.2015}. 

We use data from VAP-A, the leading probe, as it crossed the duskside plasmasphere at low $L$-shells at nearly the same time MMS crossed the duskside plasmaspheric plume at high $L$-shells. During this time, VAP-A and MMS remained within roughly 5 hours of each other in magnetic local time (MLT), as can be seen in Figure \ref{TEC}. We choose to use data from MMS-1, though the large-scale thermal and density structures that we examine look essentially identical at each of the four spacecraft. 

VAP Helium Oxygen Proton Electron (VAP-HOPE) instrument \cite{Funsten.2013} and the MMS Hot Plasma Composition Analyzer (MMS-HPCA) \cite{Young.2016} are top hat electrostatic analyzers with time of flight sensors. HOPE and HPCA measure three-dimensional mass-per-charge-discriminated plasma ion distribution functions every 22 and 10 seconds, respectively. HOPE measures directional fluxes at 72 logarithmically spaced energy steps from 1 eV/q to 50 keV/q. HPCA measures directional fluxes at 63 energy steps logarithmically spaced between roughly 1 eV/q and 40 keV/q. HOPE measures the ion species of H+, He+, and O+ and HPCA measures H+, He+, He++, and O+. The low-energy threshold of HOPE is increased to $\sim$25 eV/q during its perigee mode, which is used below roughly $L<1.5$. During its perigee mode (roughly $L\le7$), HPCA does not make measurements. Measurements from the lowest energy channels of both instruments must be used with care, because both VAP and MMS may be positively electrically charged in the plasmasphere by upwards of 1 V \cite{Goldstein.2014,SarnoSmith.2015,SarnoSmith.2016}. The sample rate of HOPE is not spin-synced, which leads to an oscillatory ``beating'' between the directions of the oversampled portion of phase space and spacecraft motion \cite{Genestreti.2016}. We do not account for the ``beating'' in the HOPE data, as its effect on the temperature is minimal compared to the temperature variations across the full span of the plume). Unlike VAP, MMS has active spacecraft potential control (ASPOC), which, when active, limits the potential of MMS to $\leq$5 V \cite{Torkar.2016}.

Additional sources of data used in this study are (1) plasma densities derived from MMS upper hybrid wave observations and continuum radiation cut-off observations made by the MMS1 electric fields double probe data \cite{Lindqvist.2016,Ergun.2016}, (2) densities derived from the spacecraft potential of VAP-A, (3) spacecraft potential measurements from both MMS-1 and VAP-A, and (4) Total Electron Content (TEC) of the F-region ionosphere derived from GPS. Median values of the ionospheric TEC were mapped to the equator from $2^\circ\times2^\circ$ grid cells at an assumed altitude of 350 km. The mapping was performed using the T04 model of the geomagnetic field \cite{TsyganenkoandFairfield.2004}. 

As in \cite{Genestreti.2016}, we use a 1-dimensional Maxwellian fitting algorithm to determine temperatures from HOPE and HPCA omnidirectional flux data, since the bulk energy of the plasmasphere is typically below the effective low-energy threshold of HOPE and HPCA ($\geq$1 eV/q). For this study, we focus solely on the low-energy ($\leq$10-100 eV) proton component. For the fits applied to MMS-HPCA data, we do not constrain the bulk energy as it is almost within the effective energy range of HPCA. For the fits applied to VAP-HOPE data, we constrain the bulk velocity to the modeled $\vec{E}\times\vec{B}$ drift velocity in the frame of the moving spacecraft. The $\vec{E}\times\vec{B}$ velocity is approximated using a centered dipole magnetic field and a Volland-Stern potential field that is parameterized by the solar wind electric field \cite{Volland.1973,Stern.1975}. \cite{Genestreti.2016} noted that even very large (100s of percent) errors in the approximation of the magnetic field do not affect the resulting fit-determined temperature. \cite{Genestreti.2016} found that the dipole-approximated and measured magnetic field strengths differed by $\sim$25\% on average for their event. We use the Volland-Stern approximation rather than the data from VAP's dedicated set of electric field probes \cite{Wygant.2013}, as small-amplitude and quasi-static electric fields are typically difficult to determine with reasonable accuracy. Note also that $\vec{E}=-\vec{v}\times\vec{B}$ cannot be determined since the bulk velocity of the cold plasma cannot be accurately determined. We use $\chi^2$ minimization to find the best-fit values of the density and temperature (as well as their associated 2$\sigma$ uncertainties) for each time-dependent measurement of the phase space density. Prior to applying the fit, we use Poisson uncertainty to assign a confidence level in each energy-dependent phase space density value.

\section{Case study of 22 October 2015}

\begin{figure*}
	\noindent\includegraphics[width=39pc]{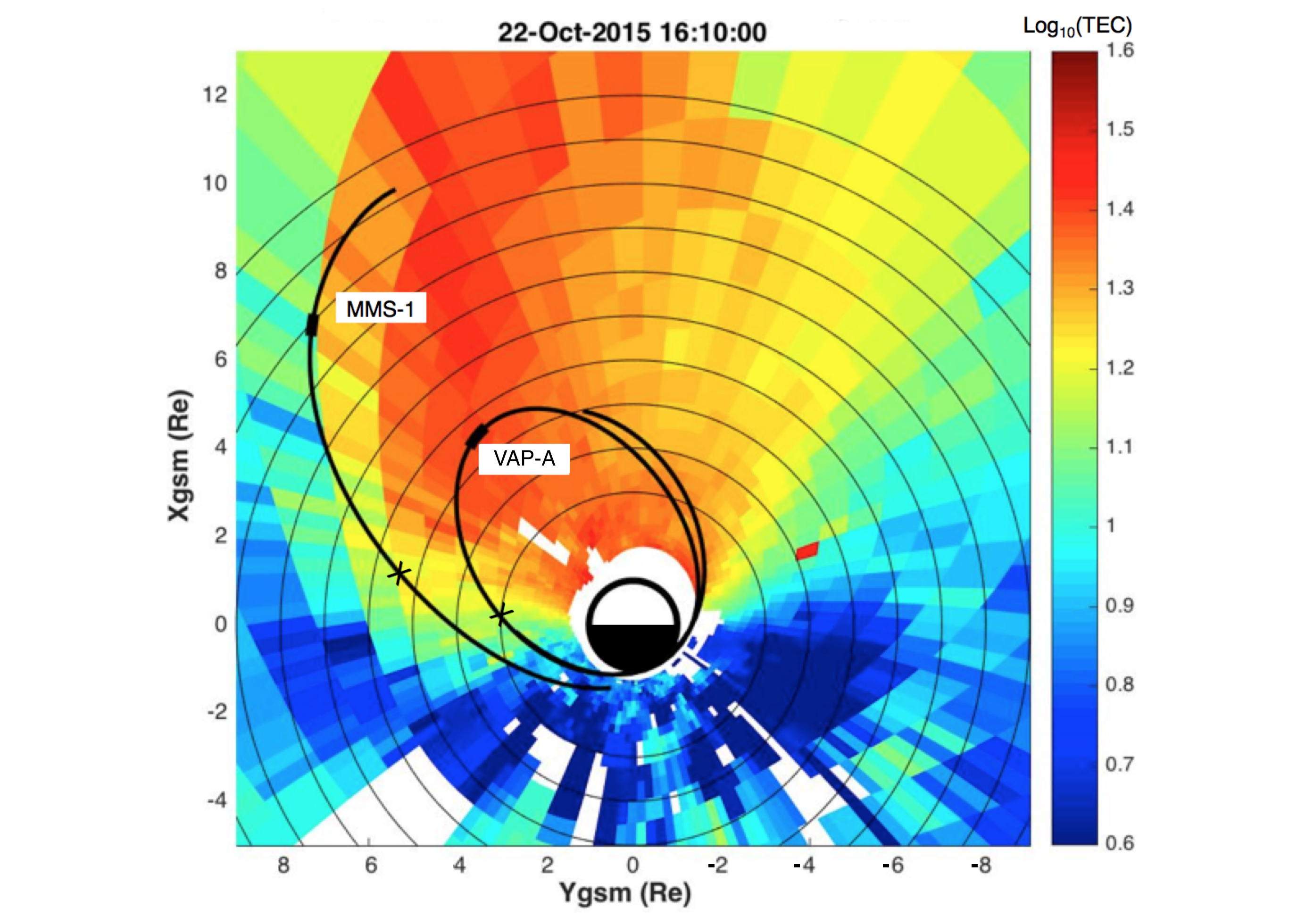}
	\caption{\label{TEC}20-minute median over 16:00--16:20 UT of the F-region TEC, which has been mapped to the equatorial magnetosphere using the T04 model. Black boxes indicate the locations of MMS-1 and VAP-A at 16:00 UT. `X'-marks indicate the locations of MMS-1 and VAP-A at 20:30 UT and 17:30 UT, respectively.}
\end{figure*}

Figure \ref{TEC} shows the orbits of MMS-1 and VAP-A for 12 hours following 22 October 2015 16:00 UT, as well as the equatorial-projected ionospheric TEC, which may be used to identify plasmaspheric plumes \cite{Foster.2002,Walsh.2014}. The TEC was calculated over the 20-minute interval 16:00--16:20 UT. After this period, differential recombination at conjugate points in the northern and southern ionospheres prevented one-to-one mapping of the ionospheric TEC to the equator. Based upon this single TEC map, it appears that MMS skirted the duskward edge of a much larger plume for 4 hours from 16:30 to 20:30 UT. This is consistent with the observations of MMS, which show large densities (1/cc $\le$ $n_{ion}$ $\le$ 10/cc) of low energy $(f_{H+}(E\le100 \mathrm{eV})\gg f_{H+}(E>100 \mathrm{eV}))$, proton-dominated ($n_{H+}/n_{tot}\sim$91\%) plasma, with a minor constituent of singly-ionized helium and negligible fluxes of low-energy oxygen and alpha particles. These are roughly consistent with (within a factor of 2 of) the average properties of the plume that were statistically determined from THEMIS data \cite{Lee.2014}. Moderate geomagnetic activity was observed throughout the conjunction and the Kp index, which is inversely correlated with the plasmasphere temperature \cite{Comfort.1986}, remained at or higher than 6. A sharp 500 nT enhancement in the auroral electrojet also occurred at roughly 18:45-19:15 UT.

For four hours prior to its entry into the magnetosphere proper (roughly 14:30 to 16:20 UT), MMS entered, exited, and reentered what was very likely a reconnection boundary layer at the magnetopause. This is evidenced by a long duration of simultaneous observations of sparse and hot magnetospheric-like ions, dense and cold ionospheric-like ions, and dense and warm magnetosheath-like protons and alpha particles. The maximum shear model \cite{Trattner.2012} predicted that MMS was within 1 $R_\mathrm{E}$ of the dayside reconnection site near 16:20 UT.

\begin{figure*}
	\noindent\includegraphics[width=40pc]{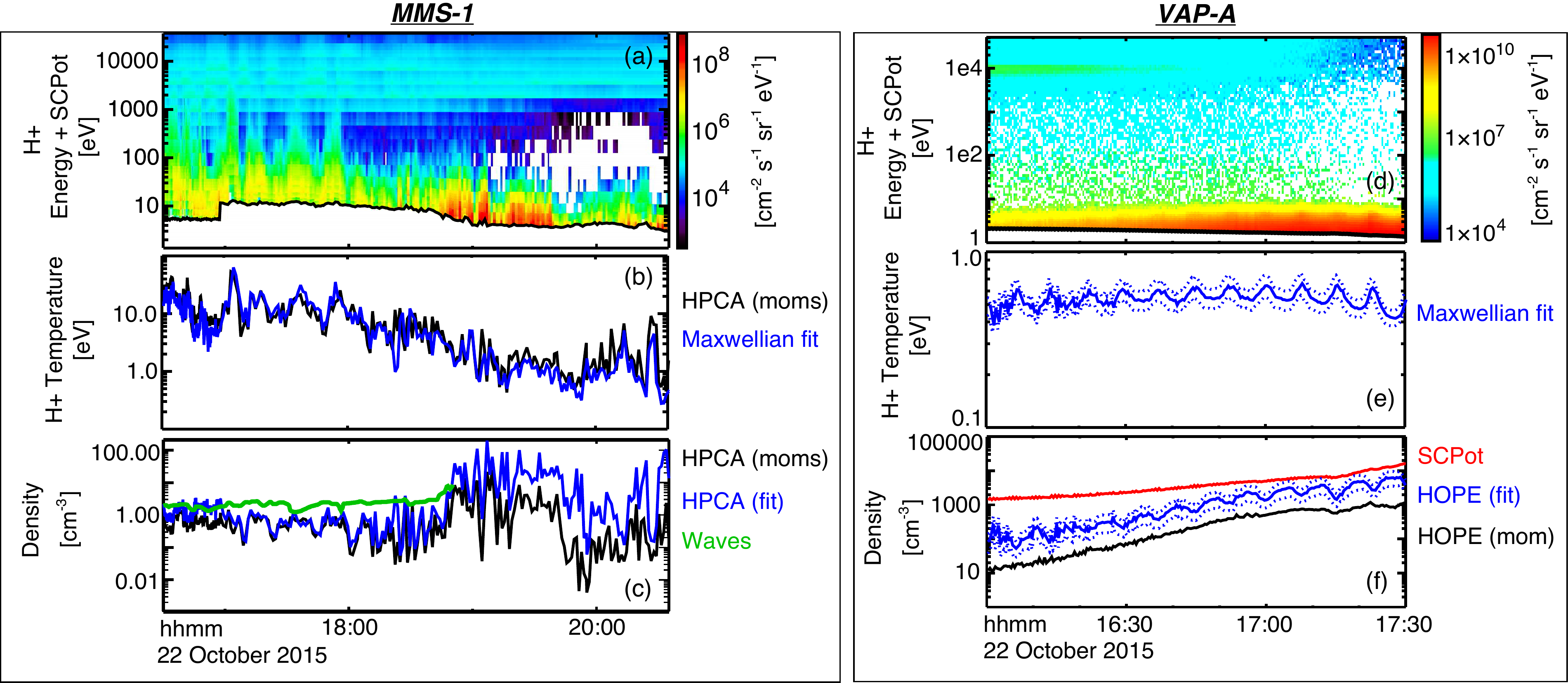}
	\caption{\label{fit_tseries}Overview of the 22 October 2015 conjunction event, where MMS and VAP provided simultaneous and nearly continuous coverage of the duskside plasmasphere and plume. Observations of the plume from MMS-1 are shown to the left and VAP-A observations of the plasmasphere main torus are shown to the right. Proton energy-intensity-time spectrograms are shown on top, where the energies have been shifted upwards by the time-dependent electric potential of the spacecraft. Standard $\pm2\sigma$ uncertainties in the fit-determined densities and temperatures, which are determined during the $\chi^2$ minimization fitting process, are shown explicitly for VAP-HOPE by the dashed blue lines in panels (e) and (f). The standard $\pm2\sigma$ uncertainties are not shown for MMS-HPCA as they are very small (less than 5\%).}
\end{figure*}

The proton temperatures and densities from MMS and VAP are shown in Figure \ref{fit_tseries}. For MMS, the fit-determined temperature and the temperature from the standard moments integration (which has been calculated in the energy range $E\le$100 eV) are nearly identical, as is shown in \ref{fit_tseries}b. The fit-determined density is lower than the waves-derived density, indicating that the peak of the density profile was not fully captured by the fitting algorithm. For VAP, we do not compare the fit-determined and standard temperatures, as the density ratio indicates that the vast majority ($\ge90\%$) of the plasma was below 1 eV (see Figure \ref{fit_tseries}f). For MMS, we use the standard temperature moment as it is nearly identical to the fit-determined temperature. The 2$\sigma$ uncertainty in the HPCA fit-derived temperature ($\pm8\%$) has not been shown for this reason. The uncertainty in the HPCA fit-derived density was 4.5 cm$^{-3}$ or $\pm27\%$. For VAP, the 2$\sigma$ uncertainties for both the density and the temperature have been shown explicitly in Figures \ref{fit_tseries}e-f (median values are $2\sigma_T/T=11\%$ and $2\sigma_n/n=36\%$). As discussed in \cite{Genestreti.2016}, the fitting algorithm is reasonably successful at determining temperatures for Maxwellian-like plasmas, but very poor at determining densities, especially when only the tail of the distribution function appears in the effective energy window of the instrument.

As discussed in \cite{Genestreti.2016}, there is a near-regular oscillatory signature in the low-energy portion of the proton energy-intensity-time spectrogram from VAP-HOPE (see Figure \ref{fit_tseries}d). The peaks / troughs in the intensity correspond to an anti-alignment / alignment between (a) the portion of phase space that is over sampled by HOPE and (b) the direction of the bulk velocity of the plasma in the spacecraft frame. This beating is strongly pronounced in the fit-determined densities and temperatures (Figures \ref{fit_tseries}e-f), causing a peak-to-median difference in the temperature of 20\%. The amplitude of these oscillations increases near perigee as the spacecraft speed increases \cite{Genestreti.2016}. (For the 22 October event, VAP-A reaches perigee after 17:30). Though this effect should certainly be accounted for in any study of smaller-scale thermal structures within the plasmasphere main torus, we find that this $\pm20\%$ deviation is nearly inconsequential compared to the variations in the temperature that are observed across the full extent of the plasmasphere and plume.

\begin{figure*}
	\noindent\includegraphics[width=33pc]{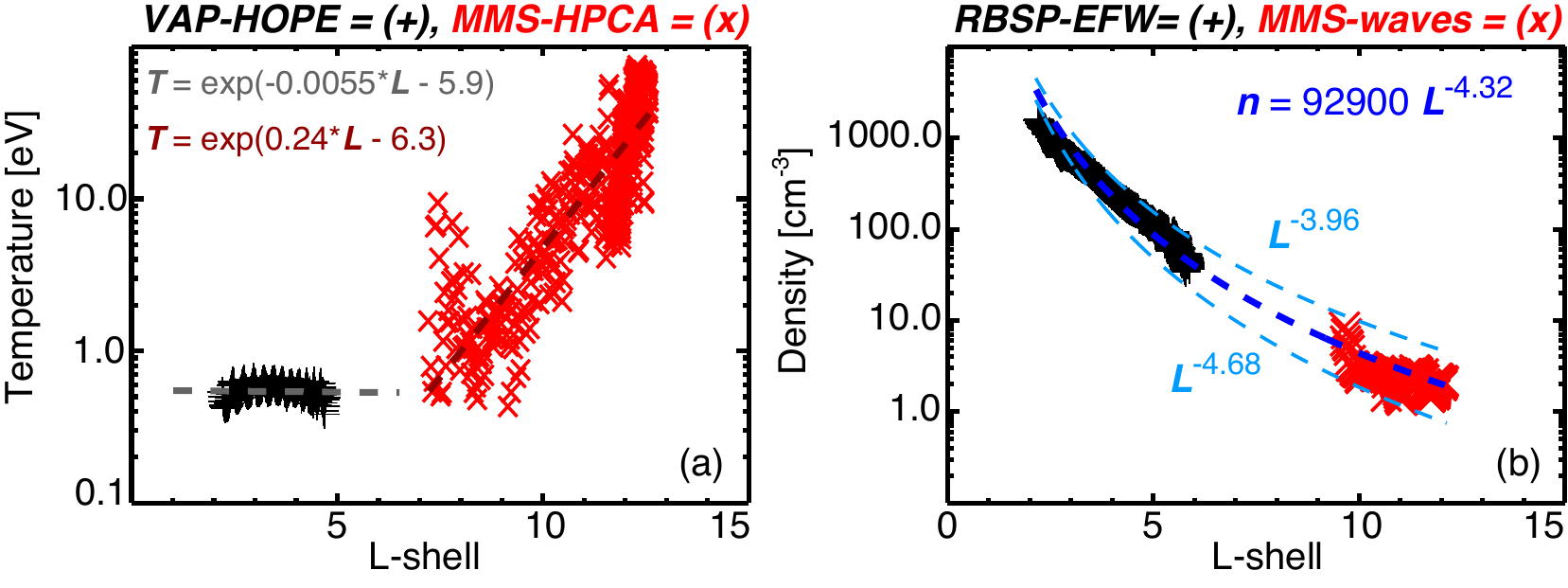}
	\caption{\label{tnlscatter}Scaling relations between the density (from MMS-EDP and VAP-EFW), temperature (from MMS-HPCA and VAP-HOPE), and $L$-shell. Fit relations are shown and listed in dark red, grey, and blue. The errors for the fit parameters are described in the text. The two light-blue dashed lines in (b) show the steepest and shallowest curves within $\pm2\sigma$ error of the best-fit curve.}
\end{figure*}

Figure \ref{tnlscatter} shows the relationships between the temperature, density, and $L$-shell locations of MMS-1 and VAP-A. MMS-HPCA temperatures from 16:30 -- 20:30 UT are from the standard moments moments integration (over $E\leq100$ eV), rather than the fit derived temperatures, though the two temperatures were nearly identical. For VAP, temperatures were derived from Maxwellian fits to the HOPE time-dependent fluxes between 16:00 -- 17:30 UT. After 17:30 UT, VAP-HOPE entered its perigee mode and the low-energy threshold was increased to 25 eV. Before 16:00 UT (near apogee), the ram energy of the spacecraft was small and the spacecraft potential was large compared to the low-energy threshold of HOPE, such that the bulk of the plasmaspheric protons were not observed. As such, it was not possible to extract reliable temperatures from the HOPE data. For MMS, waves-derived densities were calculated from 16:30 -- 18:45 UT. After this point, the upper hybrid line grew to frequencies that could not be measured by the electric field double probes. 

We have computed rough exponential scaling relations for the proton temperature (Figure \ref{tnlscatter}a) and density vs $L$-shell (Figure \ref{tnlscatter}b). First, we determined separate best fit lines for the scaling between the temperature and $L$ within the orbits of VAP and MMS. The factor of proportionality for VAP is --5.9$\pm$0.2; for MMS, it is --6.3$\pm$0.4. The exponential growth rate for MMS was 0.24$\pm$0.02, which was significantly larger than the exponential growth/decay rate for VAP, which was --0.006$\pm$0.01. An apparent ``knee'' existed, for this event at least, near geosynchronous orbit, where the temperature as a function of $L$ changed from being largely constant to highly variable. For the scaling between the density and $L$-shell, we found a multiplicative scaling factor of 93000 ($\pm2\%$ uncertainty) and a decay rate of --4.3 ($\pm0.4$). This is within error bars of the theoretical $L^{-4}$ scaling rate derived by \cite{Chappell.1974}. 

As is shown in Figure \ref{tnlscatter}, the properties of the plasmaspheric protons change dramatically from the base of the equatorial plasmasphere, near the inner radiation belt, to its outermost extent, the magnetopause. For this event, we observed a density decrease of $\sim1000\times$ and a temperature increase of $\sim1000\times$. The temperature gradient is almost exclusively observed in the plume, which was observed by MMS. In the plume, there was a significant population of higher-energy ($\sim$10 keV) protons observed along with the cold plasmaspheric protons, which may indicate that some form of cross-population interaction (e.g., collisional heating, wave-plasma interactions) is responsible for heating the plume \cite{Gallagher.2016}.

\section{Conclusions}

We established scaling relations for the temperature, density, and $L$-shell for one event, 22 October 2015, when MMS and VAP covered the entire equatorial plasmasphere and plume in a very nearly spatially continuous manner. An equatorial projection of the ionospheric total electron content (TEC) suggested that MMS and VAP skirted the duskward edge of the plasmasphere and plume. To calculate the proton temperature, we applied the 1-d Maxwellian fitting scheme of \cite{Genestreti.2016} to the VAP-HOPE and MMS-HPCA distribution function data. For MMS, the estimated and measured temperatures are nearly identical, as the plume is sufficiently accelerated and heated to appear within the energy window of MMS. Earlier in the same 22 October orbit of MMS, \cite{ToledoRedondo.2016b} identified a unique diffusive scale size intermediate between the hot magnetospheric ions and the electrons, which they associated with the presence of these cold ions. They also identified unique heating and acceleration mechanisms that occurred within this cold ion diffusion region. Later in the orbit of MMS, we determined that these cold ions, which may affect reconnection, were heated by a factor of 100 and the density was depleted by a factor of 1000 before even reaching the magnetopause. We found that the proton density scales as $L^{-4.3\pm0.4}$, which is very nearly identical to (within error bars of) the theoretical $L^{-4}$ scaling derived in \cite{Chappell.1974}.  Lastly, we noted that the vast majority of the proton heating appeared in the plume at large $L$-shells ($L>7$). In the future, it would be desirable to examine additional conjunctions between VAP and MMS so any dependence of the temperature of the plasmasphere on time, MLT, geomagnetic activity, etc. can be examined separately from its dependence on $L$, which was the focus of this study.


%
%
%
%
%
%
%

\begin{acknowledgments}
MMS data was obtained from the science data center at the University of Colorado (https://lasp.colorado.edu/mms/sdc/) and VAP electric potential data are available from the electric field and waves data repository (http://www.space.umn.edu/VAPefw-data/). The VAP-HOPE data are available from the energetic particle, composition, and thermal plasma suite data repository (https://www.rbsp-ect.lanl.gov/). The ionospheric total electron content (TEC) data were provided by Dr. John Foster in a private communication. KJG was supported by NASA contract NNG04EB99C. RN was supported by Austrian Science Funds (FWF) I2016-N20. KJG would like to thank J. Furman and J. Mukherjee for help with the HPCA data and C. Mouikis, L. M. Kistler, R. Skoug, and B. A. Larsen for help with the HOPE data.
\end{acknowledgments}

\end{article}
%
%
%
%
%
%
%
%


\end{document}